\documentclass{llncs}
\usepackage{graphicx}
\usepackage{multirow}
\begin{document}
\newcommand{\tabincell}[2]{\begin{tabular}{@{}#1@{}}#2\end{tabular}}

\title{Muscle Fatigue Analysis Using OpenSim}
\titlerunning{HCI 2017}  % abbreviated title (for running head)
\author{Jing Chang\inst{1} \and Damien Chablat\inst{1} \and 
Fouad Bennis\inst{1} \and Liang Ma\inst{2}}
\authorrunning{Jing Chang et al.} % abbreviated author list (for running head)
\tocauthor{}
\institute{Laboratoire des Sciences du Num\'erique de Nantes (LS2N), \'Ecole Centrale de Nantes, 44321 Cedex 3, Nantes, France\\
\and
Department of Industrial Engineering, Tsinghua University,\\
Beijing, 100084, P.R.China\\
\email{\{Jing.chang, Damien.Chablat, Fouad.Bennis\}@ls2n.fr,\\
liangma@tsinghua.edu.cn}}

\maketitle              % typeset the title of the contribution

\begin{abstract}
In this research, attempts are made to conduct concrete muscle fatigue analysis of arbitrary motions on OpenSim, a digital human modeling platform. A plug-in is written on the base of a muscle fatigue model, which makes it possible to calculate the decline of force-output capability of each muscle along time. The plug-in is tested on a three-dimensional, 29 degree-of-freedom human model. Motion data is obtained by motion capturing during an arbitrary running at a speed of 3.96 m/s. Ten muscles are selected for concrete analysis. As a result, the force-output capability of these muscles reduced to 60\% - 70\% after 10 minutes' running, on a general basis. Erector spinae, which loses 39.2\% of its maximal capability, is found to be more fatigue-exposed than the others. The influence of subject attributes (fatigability) is evaluated and discussed. 
\keywords{muscle fatigue analysis; digital human modeling; OpenSim; muscle fatigue model; muscle fatigability.}
\end{abstract}

%%%%---------%%%%
\section{Introduction: muscle fatigue and digital human modeling}
%%%%---------%%%%
%
\subsection{Muscle fatigue}
Muscle fatigue is defined as the decrease in maximum force \cite{Bigland-Ritchie1995}. Work-related muscle fatigue contributes to occupational Musculoskeletal Disorders (MSDs) \cite{Chaffin1999}, which makes up the vast proposition of the occupational diseases \cite{Eurogip2015}. As illustrated by Armstrong \cite{Armstrong1993}, improper physical work requirements lead to muscle fatigue. It is important to quantify fatigue and to determine the limits of acceptable work requirements \cite{Chaffin1999}. Proper work design would reduce the risk of excessive physical workload.

In the effort to involve muscle fatigue analysis into work design, there are two key problems that bother us. First, in actual working scene, the motion adopted by workers to finish a task would be arbitrary rather than routine and repeated. This makes it difficult to evaluate the exact workload carried by a certain muscle. Fatigue analysis, without the exact information about muscle workload, would be inaccurate. Second, muscle fatigue process varies a lot among human groups. The utilization of fatigue analysis would be limited without proper consideration about demographical human attribute.    
\subsection{Digital human modeling}
Digital human modeling (DHM) technique offers an efficient way to simulate ergonomics issues in the process of work design. The integration of biomechanical models with DHM systems allows us to evaluate musculoskeletal workload in manual work simulations. Related software such as Jack \cite{Jack2017}, Delmia \cite{Delmia2017}, 3DSSPP \cite{3DSSPP1999}, Anybody \cite{Damsgaard2006}, OpenSim \cite{Delp2007} are available for work design. All these softwares render realistic mannequins to visualize work tasks. Backward or inverse dynamics methods are used to calculate the muscle-tendon reaction force \cite{Chaffin2008}. 

Among the mentioned software, the mannequin used by Jack, Delmia and 3DSSPP shall be settled by the gender and the percentile of body height and weight in a given anthropometric database (USA, Canadian, German, Korean, etc). This makes it easy to apply analysis for a specific group of people. Unfortunately, as a vital parameter of work design, muscle force capacity is not included in the database. It is unreasonable to assume the same muscle capacity among different anthropometric groups while the other measures diverse. Further muscle analysis on the basis of this muscle capability would be low-effective. 
 
In Jack, Delmia and 3DSSPP, a motive task is simulated by the congregation of a set of static tasks with a certain posture. Each working posture is evaluated separately without considering the history of the motion. The external loads, the duration time and the frequency of the posture is identified. By applying strength models or inverse dynamic models, load of several major joints are calculated. In 3DSSPP and Jack, fatigue analysis is available based on the static joint load, task duration and frequency. This method goes well for simple and repetitive tasks. But when it comes to arbitrary task, there would be a great lack of accuracy.
 
The simulation and analysis by OpenSim and Anybody are more specified. The mannequin is constituted of concrete bones and muscles where musculoskeletal geometry is scaled and adaptive to subjects. The motions obtained by a motion capture system or computed along a simulated task permitted us to have the kinematic information, such as positions, velocities and accelerations of a motion. The inertial properties of body segments are estimated. By applying the Newtonian principles, the prediction of the resultant extrinsic forces and moments are then available.

Although the joint reaction force and muscle load are accessible in OpenSim and Anybody, the accumulation effect has not been taken into account; and no accurate fatigue analysis is available.
\subsection{Objectives}
In this research, a plug-in to OpenSim is written to involve the muscle fatigue analysis to an arbitrary task. Concrete muscle force capability change is specified and the influence of demographic human attributes is considered. This work is promising to offer a virtual work design platform that helps to predict muscle fatigue.
%%%%---------%%%%
\section{Methodology: OpenSim human modeling and muscle fatigue analysis}
%%%%---------%%%%
%
\subsection{Human modeling and dynamic simulation in OpenSim}\label{chap2.1}
As mentioned above, OpenSim is a digital human modeling platform. It allows users to build and analyze different musculoskeletal models. A model consists of a set of rigid segments connected by joints. Muscles and ligaments span the joints, develop force, and generate movements of the joints. After the build-up of a musculoskeletal model, OpenSim takes experimentally-measured kinematics, reaction forces and moments as input data. This is usually obtained by motion capture system from a subject. The experimental kinematics ($i.e.$, trajectories of markers, joint centers, and joint angles) are used to adjust and scale the musculoskeletal model to match the dimensions of the subject\cite{Delp2007}. 

For dynamic simulation, an inverse kinematics problem is solved to find the model joint angles that best reproduce the experimental kinematics. Then a residual reduction algorithm is used to adjust the kinematics so that they are more dynamically consistent with the experimental reaction forces and moments. Finally, a  computed muscle control (CMC) algorithm is used to find a set of muscle excitations and distribute forces across synergistic muscles to generate a forward dynamic simulation that closely tracks the motion \cite{Delp2007}. In this way, the workload of each muscle is accessible along an arbitrary motion, which paves way for the fatigue analysis. 
\subsection{Muscle fatigue analysis}\label{chap2.2}
 Ma et al. \cite{Ma2009} proposed a "Force-load fatigue model” based on mechanical parameters. It depicts how muscle force declines with time with consideration of relative workload and intrinsic human attribute. The model was described as a differential function (Eq. \ref{eq}). According to this model, during a fatiguing process, muscle force capability ($F_{cem}(t)$) changes depending on a) Maximal (or initial) muscle force capability, $F_{max}$; b) External load on the muscle, $F_{Load}(t)$ and c) Intrinsic muscle fatigability, $k$. For detailed explanation of this model was introduced in Ma et al. \cite{Ma2009,Ma2011} and \cite{Sakka2015} for static and dynamic cases, respectively.
 \begin{eqnarray}\label{eq}
\frac{d F_{cem}(t)}{d t}=-k\frac{F_{cem}(t)}{F_{max}}F_{Load}(t)
\end{eqnarray}

This model has been mathematical validated in Ma et al. \cite{Ma2009} with static MET models and other existing dynamic theoretical models.
 
In this model, intrinsic human attribute concerning to fatigue rate is taken into consideration, which is referred to fatigability. The definition of fatigability is proposed by Ma \& Chang \cite{Ma2014} 
``Muscle fatigability describes a tendency of a muscle from a given subject to get tired or exhausted, and it should only be determined by the physical and psychological properties of the individual subject". According to this model, the decrease rate of muscle capability is in proportion with both work load and current muscle capability. The proportion coefficient $k$ quantifies the tendency of muscle strength descending, and is noted as fatigability. 

Fatigability varies significantly among human groups. For example, females are found to be more fatigue-resistant than males \cite{Yoon2015}; and the older groups shows significantly much less force loss than the younger group after a certain exercises \cite{Kent-Braun2002}. Fatigability $k$ has been determined by comparing the Force-Load muscle fatigue model with the empirical maximal endurance models \cite{Ma2011}. The determined value of $k$ varies from 0.87 $min^{-1}$ to 2.15 $min^{-1}$ for general muscle groups. Ma et al \cite{Ma2013} also conducted experiments to measure fatigability. In a static drilling task, the fatigability of 40 male workers was identified to be $1.02 \pm 0.49$ $min^{-1}$ for the upper limbs. %CD
\subsection{Muscle fatigue analysis in OpenSim}
The object of this research is a concrete analysis of muscle fatigue. In another word, how force capability of each muscle declines during an arbitrary motion. According to the Force-load muscle fatigue model, this objective can be achieved on condition of two values: workload on each muscle along the motion and the maximal muscle capability. 
\subsubsection{Workload on each muscle}
The muscle force generation dynamics can be divided into activation dynamics and contraction dynamics \cite{Zajac1989}. Activation dynamics corresponds to the transformation of neural excitation to activation of the muscle fibers. Muscle contraction dynamics corresponds to the transformation of activation to muscle force. 

 \begin{figure}[h!]
 \label{Fig1}
 \centering
 \includegraphics[width=1\textwidth]{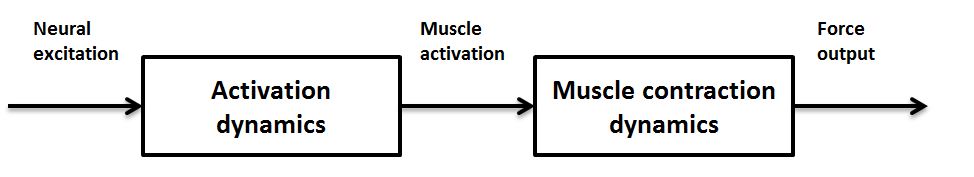}
  \caption{Muscle force generation dynamics. Adapted from Zajac (1989) \cite{Zajac1989}. }
 \end{figure}
 Activation dynamics is related to calcium release, diffusion and uptake from the sarcoplasmic reticulum \cite{Winters1995}. It is modeled by a first-order differential equation \cite{Thelen2003}:
 \begin{eqnarray}\label{eq2}
 \frac{da}{dt}=\frac{u-a}{\tau(u,a)}
 \end{eqnarray}
 where $u$ is excitation (from 0 to 1), $a$ is activation (from 0 to 1), and $\tau$ is a variable time constant.

Muscle contraction dynamics deals with the force-length-velocity relationships and the elastic properties of muscles and tendons. In OpenSim, it is modeled by a lumped-parameter model \cite{Delp2007}:

\begin{eqnarray}
\frac{dl_{m}}{dt}=f_{v}^{-1}(l_{m}, l_{mt},a)
\end{eqnarray}
where $l_{m}$ is the muscle length, $l_{mt}$ is the muscle-tendon actuator length, and $f_{v}$ is the force velocity relation for muscle.

As mentioned in chapter \ref{chap2.1}, OpenSim develops a CMC algorithm to calculate muscle activation and therefore to distribute joint force among a series of muscles. 
By applying the CMC algorithm, the workload on each muscle is accessible. 
\subsubsection{The Maximal muscle force capability of each muscle}
Muscle activation depends on the neural excitation level. In the case of a certain muscle contraction speed and muscle length, muscle force increases with muscle activation. Full activation (i.e., $a(t) = 1$) happens when a muscle contractile component has been maximally excited (i.e., $u(t) = 1$) for a long time \cite{Zajac1989}. During an arbitrary motion, muscle kinematics changes from time to time. We calculate the maximal muscle force $F_{max}$ based on the CMC algorithm, in addition that the activation level of each muscle is preset to full level ($a(t) = 1$).
 
As the final step, a Plug-in is written based on the Force-load muscle fatigue model, with the required inputs obtained by the above methods. 
%%%%---------%%%%
\section{Data and Simulation}
%%%%---------%%%%
The plug-in is tested on a three-dimensional, 29 degree-of-freedom human model developed by Stanford \cite{Hamner2010}. The model, as the other OpenSim musculoskeletal models, is made up of bodies, joints, and muscle-tendon actuators. Specifically, this model consists of 20 body segments, 19 joints and 92 muscle actuators, as shown in Figure~2. The inertial parameters for the body segments in the model are based on average anthropometric data obtained from five subjects (age 26 $\pm$ 3 years, height 177 $\pm$ 3 cm, and weight 70.1 $\pm$ 7.8 kg).

 \begin{figure}[h!]
 \label{Fig2}
 \centering
 \includegraphics[width=0.35\textwidth]{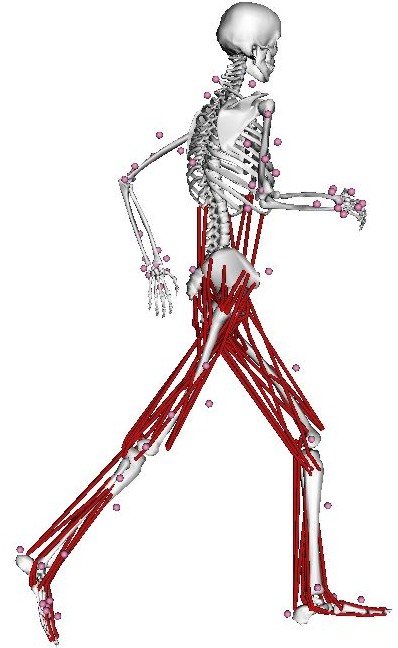}
  \caption{Full body running digital human modeling. Muscles are represented by red lines.}
 \end{figure}

This model is developed to study the muscles' contribution to the acceleration of the body during running. It covers the muscles that needed for arbitrary running motions. These muscles could be classified into three groups: torso-core muscle group, pelvis-femur muscle group, and lower knee extremities muscle group.

The simulation data is also from the project of Hamner et al. \cite{Hamner2010}. It is recorded from a healthy male subject (height 1.83 m, mass 65.9 kg) running on a treadmill at 3.96 m/s. A total of 41 reflective markers are placed on the subject’s anatomical landmarks during the experiment to scale the model to the subject’s anthropometry. Ground reaction forces and markers' trajectories are recorded. The recorded motion lasts for 10 seconds.

In our study, 10 muscles are selected from the three muscle groups to conduct muscle fatigue analysis. The basic characteristics of these muscles are listed in Table~\ref{tab1}.

\begin{table}[h!]
\label{tab1}
\begin{center}
\caption{Basic characteristics of the analyzed muscles.}
\begin{tabular}{l@{\qquad}l@{\qquad}l@{\qquad}l}
\hline
\rule{0pt}{20pt} 
Muscle name & \tabincell{l}{Appertained \\ group} & \tabincell{l}{Optimal fiber\\length ($m$)} & \tabincell{l}{Maximal isometric\\force ($N$)} \\[2pt]
\hline\rule{0pt}{12pt}
Erector spinae        & Torso-core    & 0.120 & 2500.0 \\
External oblique      & Torso-core    & 0.120 & 900.0  \\
Internal oblique      & Torso-core    & 0.100 & 900.0  \\
Adductor magnus       & Pelvis-femur  & 0.131 & 488.0  \\
Glute maximus         & Pelvis-femur  & 0.142 & 573.0  \\
Glute medius          & Pelvis-femur  & 0.065 & 653.0  \\
Tibialis posterior    & Lower knee    & 0.031 & 1588.0 \\
Lateral gastrocnemius & Lower knee    & 0.064 & 683.0  \\
Extensor digitorum    & Lower knee    & 0.102 & 512.0  \\
Soleus                & Lower knee    & 0.050 & 3549.0  \\
[2pt]
\hline
\end{tabular}
\end{center}
\end{table}

%%%%---------%%%%
\section{Results of simulation}
\begin{figure}[h!]
 \label{Fig3}
 \centering
 \includegraphics[width=0.8\textwidth]{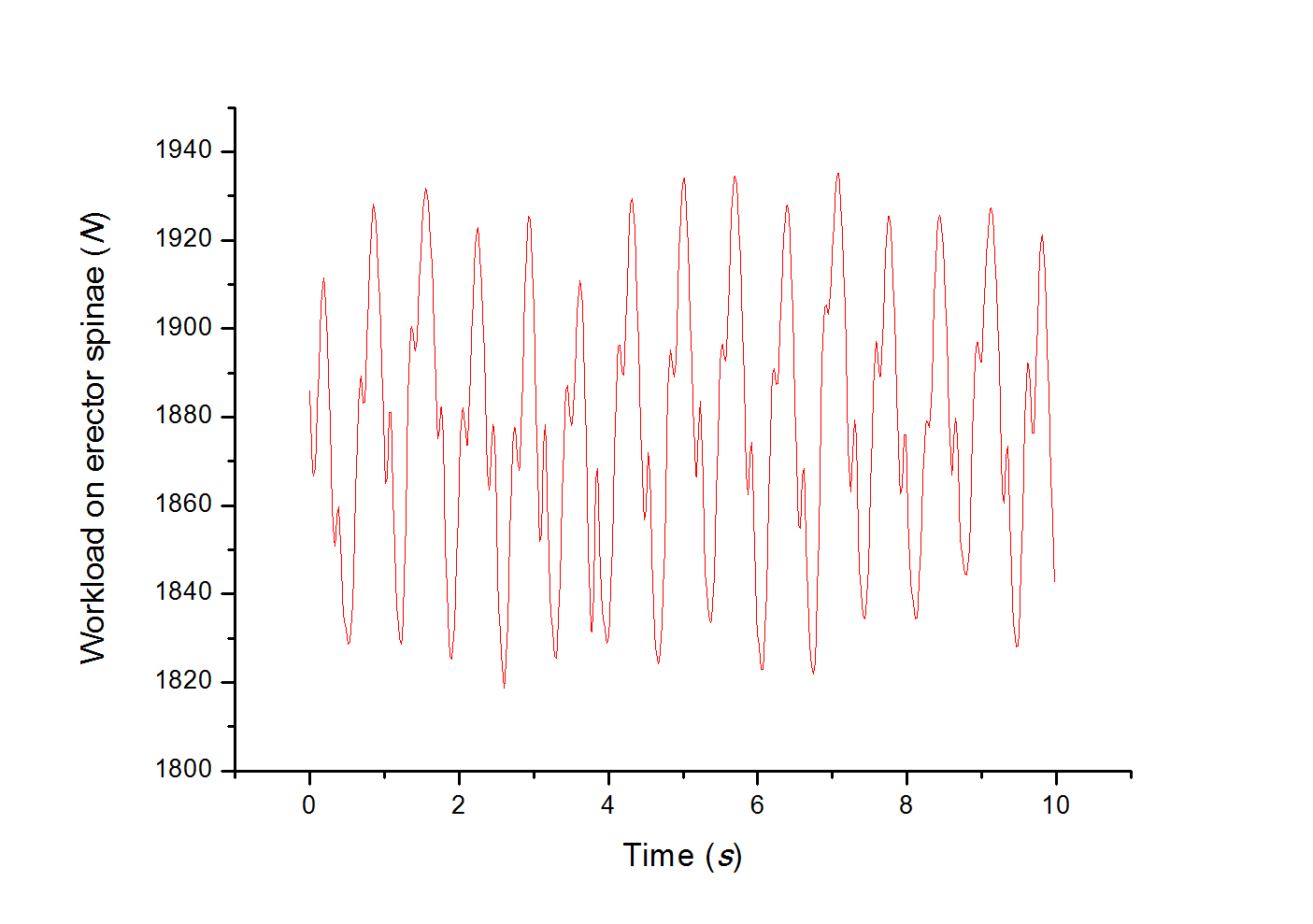}
  \caption{Workload on erector spinae muscle during 10s arbitrary running at 3.96 m/s.}
 \end{figure}

%%%%---------%%%%
During the arbitrary running, workloads on muscles vary from moment to moment. A typical muscle workload change is shown in Figure~\ref{Fig3}.

%%%%%%%%%%%%
\subsection{General muscle force decline}

In order to investigate the fatigue process, the motion data is duplicated to 10~min. During the running process, muscle force capabilities decline with time. After input the fatigability ($k$) of the subject, the detail information about force capability changes is accessible. A general view of force capability changes of the selected muscles are shown in Figure~4. Here the fatigability is set to 1.0 $min^{-1}$. Generally, the muscles' capabilities reduce to 60\% to 70\% of their maximum after running  for 10 min.

 \begin{figure}[h!]
 \label{Fig4}
\begin{centering}
 \includegraphics[width=0.7\textwidth]{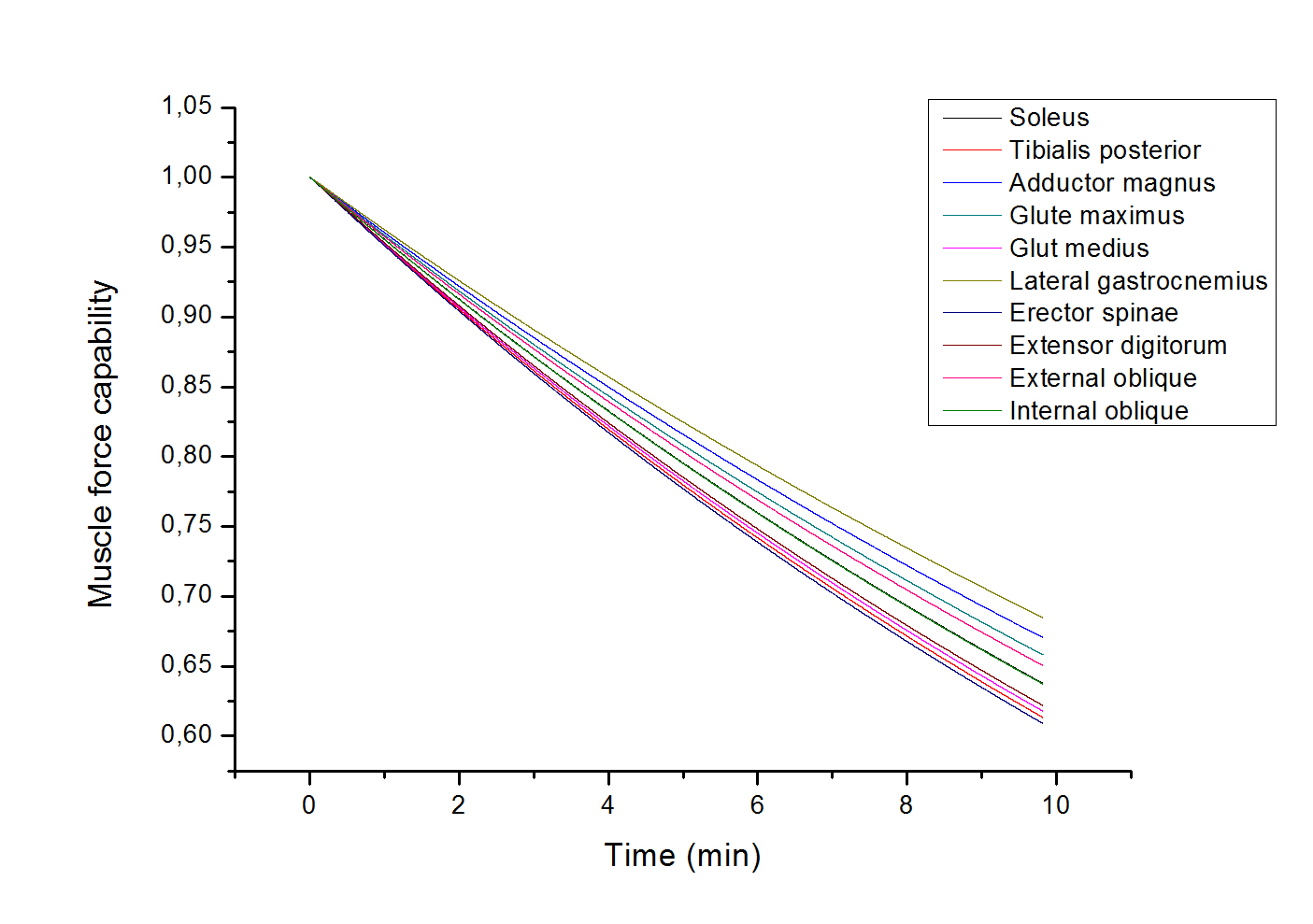}
  \caption{Force capability lines of ten muscles during 10 min running. $k$ = 1.0 $min^{-1}$. \qquad \qquad Force of each muscle is normalized by its maximum.}
  \end{centering}
 \end{figure}
\subsection{Comparison among different muscles}
The initial and ending force capabilities of the selected muscles are shown in Table~2.

\begin{table}[h!]
\label{tab2}
\begin{center}
\caption{General information of muscle force capabilities}
\begin{tabular}{l@{\qquad}l@{\qquad}l@{\qquad}l@{\qquad}l}
\hline
\rule{0pt}{20pt} 
Muscle name & \tabincell{l}{Appertained \\ group} & \tabincell{l}{Initial \\ capabilities\\($N$)} & \tabincell{l}{Finial \\ capabilities\\($N$)} & \tabincell{l}{Proportion\\of reduction}\\[2pt]
\hline\rule{0pt}{12pt}
Erector spinae        & Torso-core    & 38703.5 & 23563.3 & 39.1\% \\
External oblique      & Torso-core    & 19876.2 & 12926.2 & 35.0\%\\
Internal oblique      & Torso-core    & 18374.8 & 11705.9 & 36.3\%\\
Adductor magnus       & Pelvis-femur  & 10923.7 & 7322.3  & 33.0\%\\
Glute maximus         & Pelvis-femur  & 12304.1 & 8096.3  & 34.2\%\\
Glute medius          & Pelvis-femur  & 19134.0 & 11817.8 & 38.2\%\\
Tibialis posterior    & Lower knee    & 36901.9 & 22624.1 & 38.7\%\\
Lateral gastrocnemius & Lower knee    & 18231.9 & 12481.3 & 31.5\%\\
Extensor digitorum    & Lower knee    & 8940.7  & 5558.1  & 37.8\%\\
Soleus                & Lower knee    & 95221.7 & 60708.3 & 36.3\%\\
[2pt]
\hline
\end{tabular}
\end{center}
\end{table}

The proportion of force capability reduction is between 30\% to 40\% for each of the ten muscles. Erector spinae loses the maximal proportion of force. As far as the selected muscles, torso-core muscle group fatigues no less than the pelvis-femur or the lower knee group (average fatigue level: (36.8 $\pm$ 2.1)\%, (35.1 $\pm$ 2.7)\%, (36.1 $\pm$ 3.2)\%, respectively).  
\subsection{Influence of fatigability index $k$}
Fatigability is a subject-specific parameter that might also varies between muscle groups. As mentioned in Chapter~\ref{chap2.2}, the determined value of $k$ varies from 0.87 $min^{-1}$ to 2.15 $min^{-1}$ for general muscle groups. In the current study, we examine the influence of fatigability by comparing the fatigue level when $k$ = 1.0 $min^{-1}$ with that when $k$ = 2.0 $min^{-1}$. A typical comparison is shown in Figure~5.

 \begin{figure}[h!]
 \label{Fig5}
 \centering
 \includegraphics[width=0.8\textwidth]{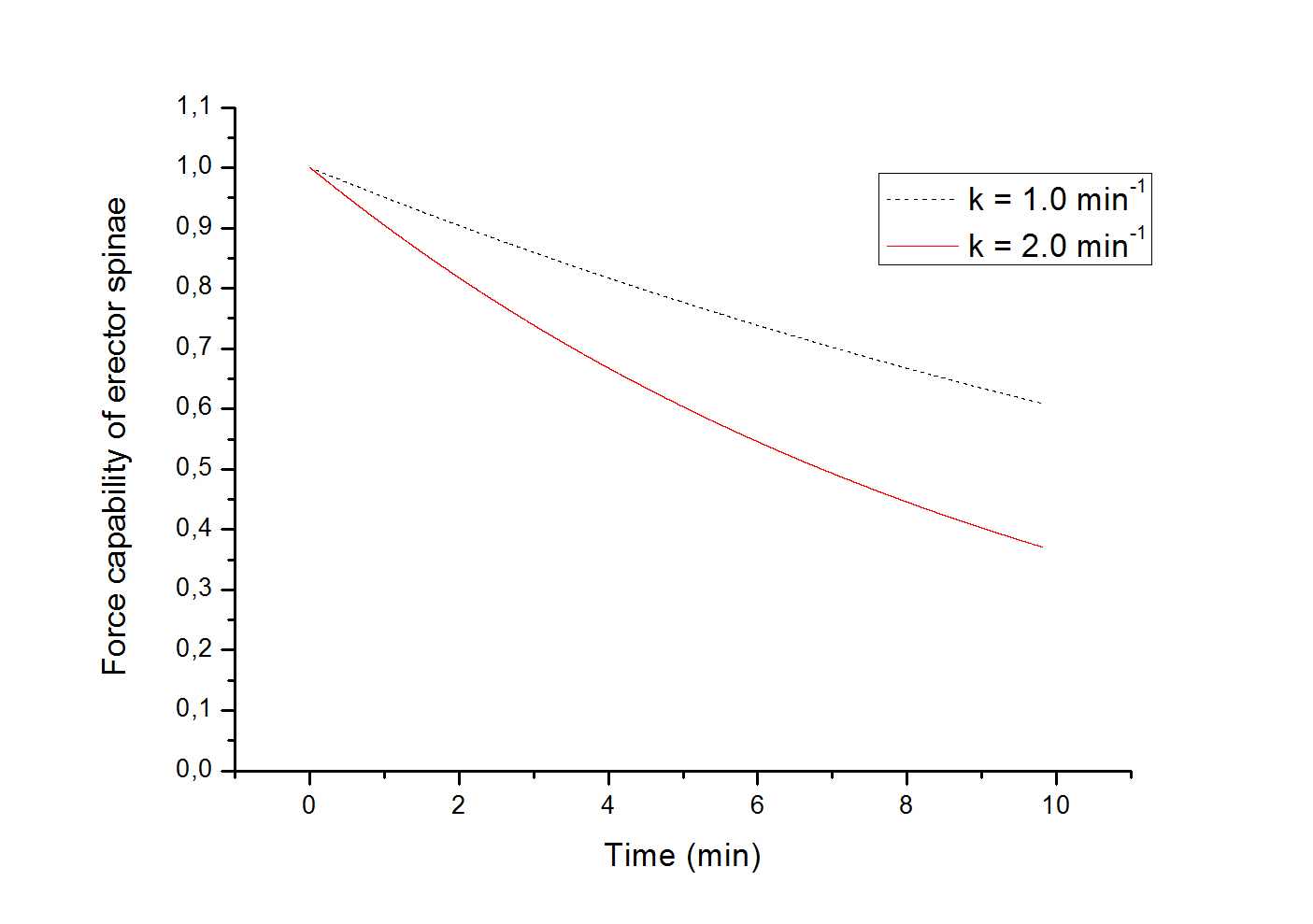}
  \caption{Force capability lines of erector spinae with different preset $k$ values. Force is normalized by the maximum.}
 \end{figure}

Table~3 manifests the fatigue level comparisons of all the ten muscles. Generally, the muscle reduces to 40\% to 50\% of its initial maximal capability. The relative sort of muscles' fatigue level remains unchanged.  

 \begin{table}[h!]
 \label{tab3}
 \begin{center}
 \caption{Comparison of muscle force capabilities between different fatigabilities}
 \begin{tabular}{l@{\qquad}l@{\qquad}p{2.2cm}@{\qquad}c}
 \hline
 \rule{0pt}{15pt} 
 ~Muscle name & 
 ~Appertained group~ &
 \multicolumn{2}{c}{~Proportion of capability loss~}\\
\cline{3-4} 
 \rule{0pt}{10pt} 
 & &  \centering $k$ = 1.0 $min^{-1}$ & $k$ = 2.0 $min^{-1}$\\  
 \hline
 \rule{0pt}{12pt} 
Erector spinae        & Torso-core    &  \centering 39.1\%  &  62.9\% \\
 External oblique      & Torso-core    &  \centering 35.0\%  &  57.7\% \\
 Internal oblique      & Torso-core    &  \centering 36.3\%  &  59.4\% \\
 Adductor magnus       & Pelvis-femur  &  \centering 33.0\%  &  55.1\% \\
 Glute maximus         & Pelvis-femur  &  \centering 34.2\%  &  56.7\% \\
 Glute medius          & Pelvis-femur  &  \centering 38.2\%  &  61.9\% \\
 Tibialis posterior    & Lower knee    &  \centering 38.7\%  &  62.4\% \\
 Lateral gastrocnemius & Lower knee    &  \centering 31.5\%  &  53.1\% \\
 Extensor digitorum    & Lower knee    &  \centering 37.8\%  &  61.4\% \\
 Soleus                & Lower knee    &  \centering 36.3\%  &  59.4\% \\
 \hline
 \end{tabular}
 \end{center}
 \end{table}
%%%%---------%%%%
\section{Discussions}
%%%%---------%%%%
According to the simulation data, after 10 minutes' running at the speed of 3.96 $m/s$, a healthy male subject is likely to lose 30\% to 40\% of his maximal muscle capability. The requirement of the running task at the current posture is about 5\% of his maximal muscle capability. If the task continues, there would be a time in future when the subject's muscle capability reduces to near to the required workload. Muscles will enter into a risk zone \cite{Ma.R2012,Ma2009a}. Damage will occur to muscles, which increases the risk of MSDs. The subject would change his posture unconsciously \cite{fuller2009}. It is important to predict the exhausting time in the early process of work design.

In Hammer's research \cite{Hamner2010}, the quadriceps (pelvis-femur muscle group) and plantar flexors (lower knee muscle group) are the major contributors to acceleration of the body mass center during running, compared with the torso-core muscle group (erector spinae and iliopsoas). While in the current study, erector spinae, who loses 39.2\% of its maximal capability, is found to be more fatigue-exposed than the others. Also, torso-core muscle group fatigues no less than the other two muscle groups. This phenomenon indicates that torso-core muscles undertake other supportive functions than contributing to body accelerating, such as counterbalancing the vertical angular momentum of the legs.

A subject with a higher fatigability value ($k$ = 2.0 $min^{-1}$) losses about 20\% of his maximal capability more than the subject with a lower fatigability value ($k$ = 1.0 $min^{-1}$). The influence of fatigability is evident. It is essential to study the fatigability among different human groups.

It should be noticed that the current study considers no effect of fatigue recovery. Future study should integrate the muscle recovery for more accurate calculation.
%%%%---------%%%%
\section{Conclusions and implications}
%%%%---------%%%%
In this study, a plug-in to OpenSim is written on the base of the Force-load muscle fatigue model and muscle force generation dynamics to obtain the concrete information about how force-output capability of each muscle declines along time. 

Simulation on a three-dimensional, 29 degree-of-freedom human model shows that the force-output capability of ten selected muscles reduced to 60\% - 70\% after 10 minutes' running at the speed of 3.96 $m/s$, with a fatigability value of 1.0 $min^{-1}$. Torso-core muscle group, which has been found to contribute less to the body's acceleration in previous research, shows no less proportion of force loss than the other two groups. The difference in fatigue level caused by the change of fatigability is evident, which emphasizes the necessity of the study and determination of fatigability among different human groups.

This work offers a virtual work design platform that helps to predict muscle fatigue and thereby to control the MSDs risks at the early stage of work design. In future works, the Motion Capture System of the \'Ecole Centrale de Nantes will be used to acquire data for industrial tasks. A force platform will be used to validate the muscle properties during experiments. The study of the fatigability $k$ as a function of the people will also be addressed in these new experiments.
%%%%%%%%%%%%%
\section{Acknowledgments}
%%%%%%%%%
This work was supported by the National Natural Science Foundation of China under Grant numbers 71471095, by Chinese State Scholarship Fund, and by INTERWEAVE Project (Erasmus Mundus Partnership Asia-Europe)  under Grants number  IW14AC0456 and IW14AC0148.

\end{document}